\documentclass[
  ,draft            
  ,numberedheadings 
  ]
  {aipproc}

\layoutstyle{6x9}

\begin{document}

\title[Using the Balance Function to study the charge correlations of hadrons]{Using the Balance Function to study the charge correlations of hadrons}


\author{P. Christakoglou, A. Petridis, M. Vassiliou for the NA49 and ALICE collaborations}{
  address={Physics Department - University of Athens - 15771 - Athens, Greece}
}

\vspace{-1.5 cm}

\begin{abstract}

We present the recent Balance Function (BF) results obtained by the NA49 collaboration for the pseudo-rapidity dependence of non-identified charged particle correlations for two SPS energies. Experimental results indicate a clear centrality dependence only in the mid-rapidity region. The results of an energy dependence study of the BF throughout the whole SPS energy range will also be discussed. In addition, the correlation of identified hadrons is studied and presented for the first time. The study of hadron correlation has also been extended in order to cope with the high multiplicity environment that is expected to be seen at LHC. We will present the latest results from simulations concerning the extension of these studies to the ALICE experiment.

\end{abstract}

\maketitle


\vspace{-0.5 cm}

The study of correlations and fluctuations on an event by event basis is expected to provide additional information on the reaction mechanism of high energy nuclear collisions \cite{QM04}. The Balance Function (BF), introduced by Bass, Danielewicz and Pratt \cite{Pratt}, provides the means to explore the space-time evolution of the emitted hadrons after such a collision. The method was initially proposed \cite{Pratt} to be related to the time of hadronization: if a pair of oppositely charged particles was produced late in the reaction, then the particles are tightly correlated in space and their relative momenta can be determined by the breakup temperature. On the other hand, if such a pair is produced at an early stage of the reaction, then the particles may diffuse apart from one another. The previous two opposite mechanisms are reflected in the BF's width and result in a narrow distribution for the late stage hadronization and a wide one for the early stage hadron production \cite{STAR_BF,NA49_BF}.


\vspace{0.1 cm}

\begin{figure}
\includegraphics[height=.22\textheight,width=.35\textwidth]{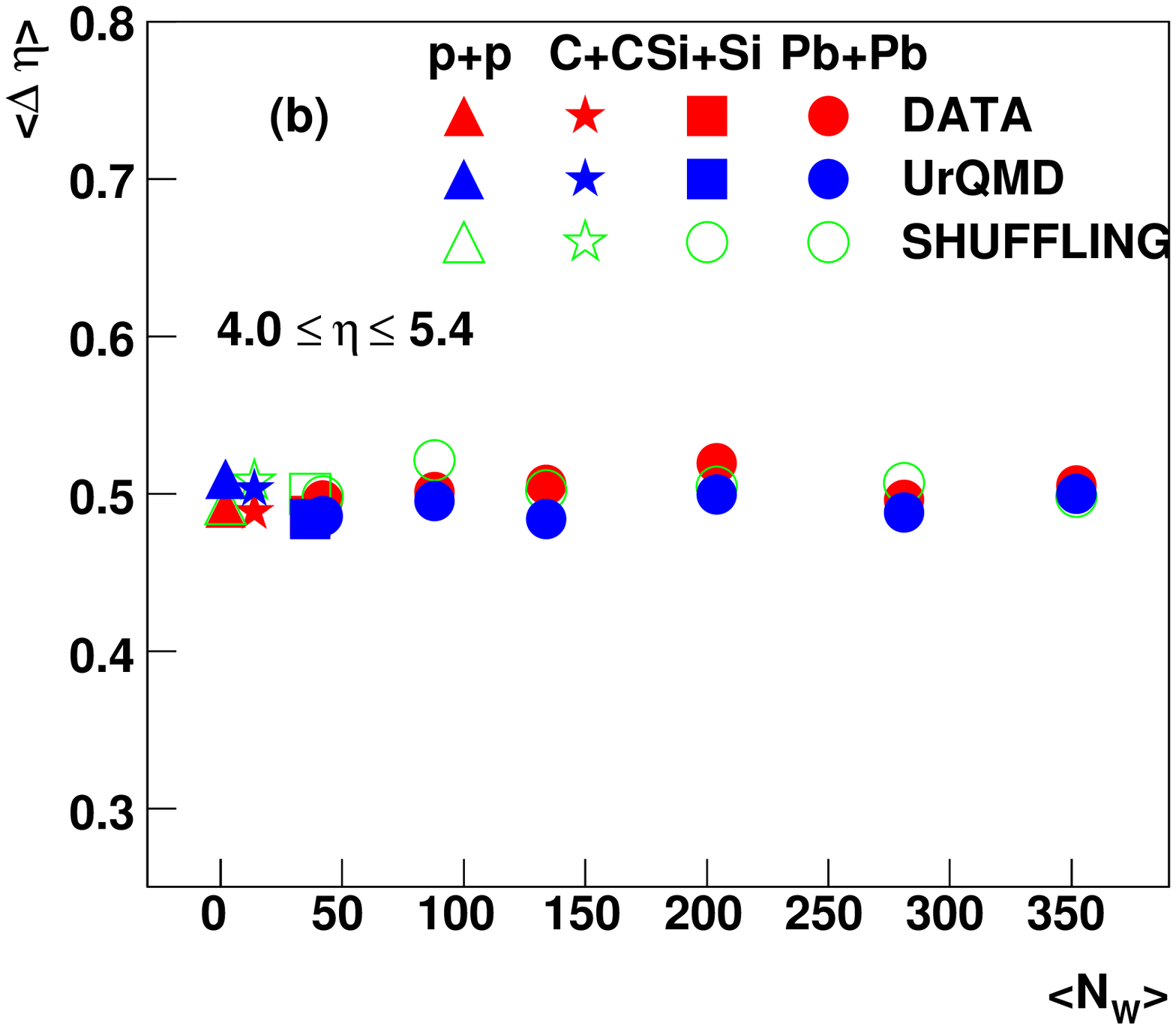}
\includegraphics[height=.22\textheight,width=.35\textwidth]{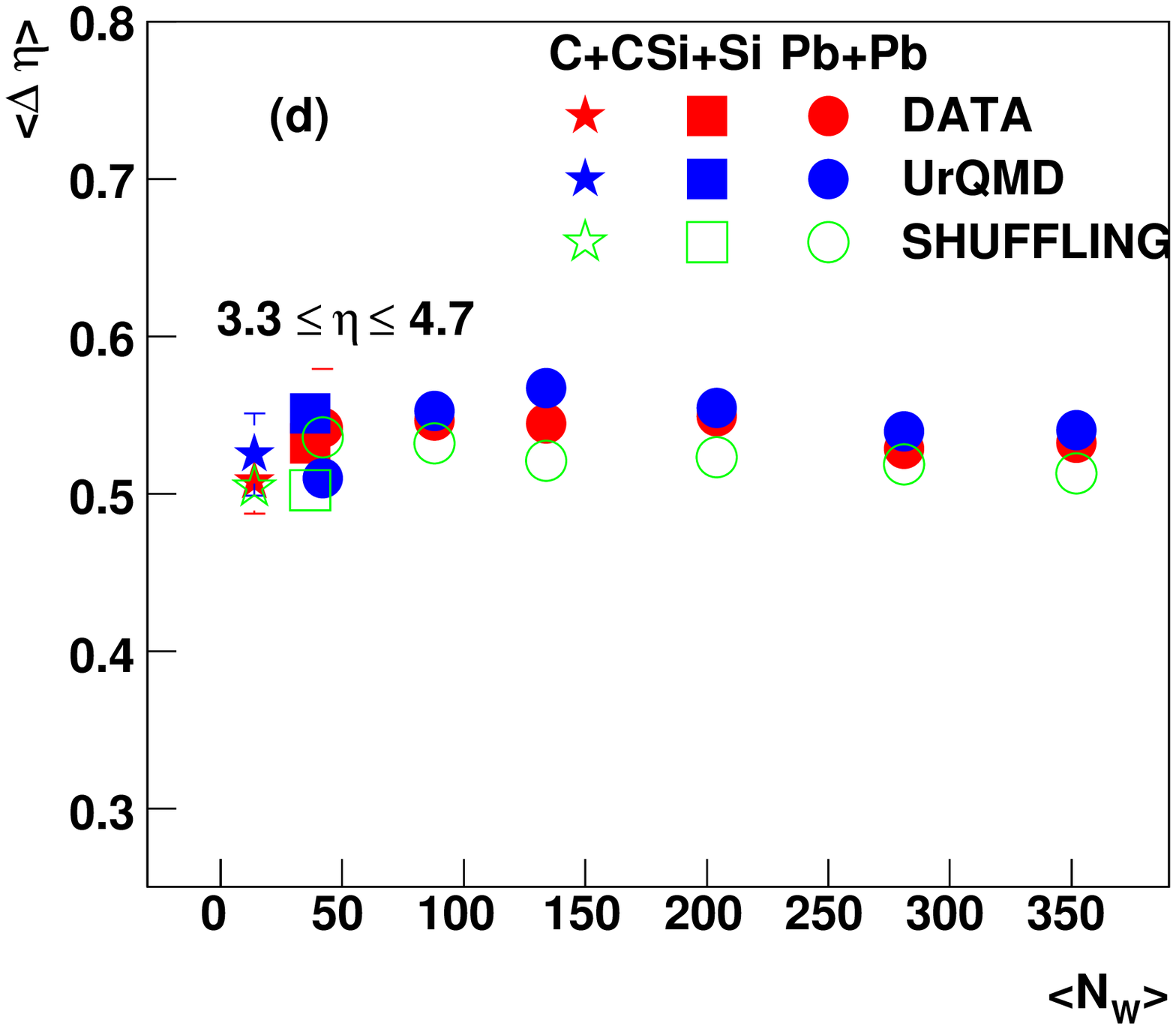}
\includegraphics[height=.22\textheight,width=.35\textwidth]{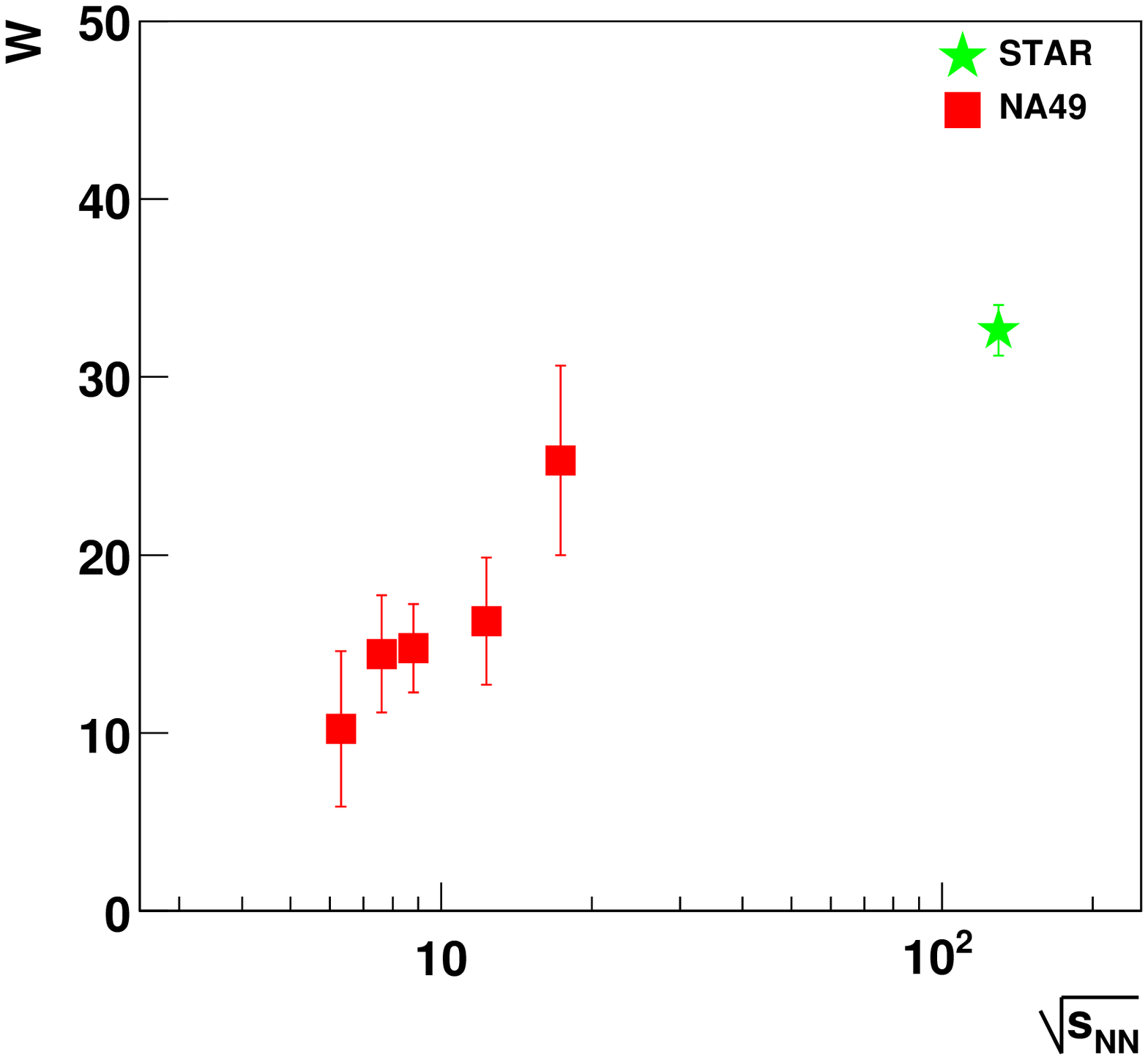}
\caption{The system size and centrality dependence of the measured width of the BF for charged particles at $\sqrt{s_{NN}} = 17.3$ GeV (plots a and b) and $\sqrt{s_{NN}} = 8.8$ GeV (plots c and d) as a function of $\langle N_{W} \rangle$ for the two different rapidity intervals. The two right plots show the dependence of the normalized parameter W, which indicates the relative decrease of the width of the BF between experimental and shuffled data, on $\sqrt{s_{NN}}$ for central Pb+Pb collisions in the SPS energy range (upper plot) and its evolution towards higher RHIC energies (lower plot).}
\label{FigNA49RapidityDependence}
\end{figure}

The BF was studied by the NA49 collaboration in order to investigate the system size and the centrality dependence of the width at two SPS energies ($\sqrt{s} = 17.3$ GeV and $\sqrt{s} = 8.8$ GeV) and in two different rapidity intervals (mid-rapidity and forward rapidity region). Fig. \ref{FigNA49RapidityDependence} shows the width of the BF distributions for real, UrQMD \cite{Urqmd} and shuffled data \cite{STAR_BF,NA49_BF} as a function of the mean number of wounded nucleons, for the two rapidity regions analyzed for both energies. There is an apparent centrality dependence for experimental data only in the mid-rapidity regions (plots a and c).

\begin{figure}
\includegraphics[height=.22\textheight,width=.35\textwidth]{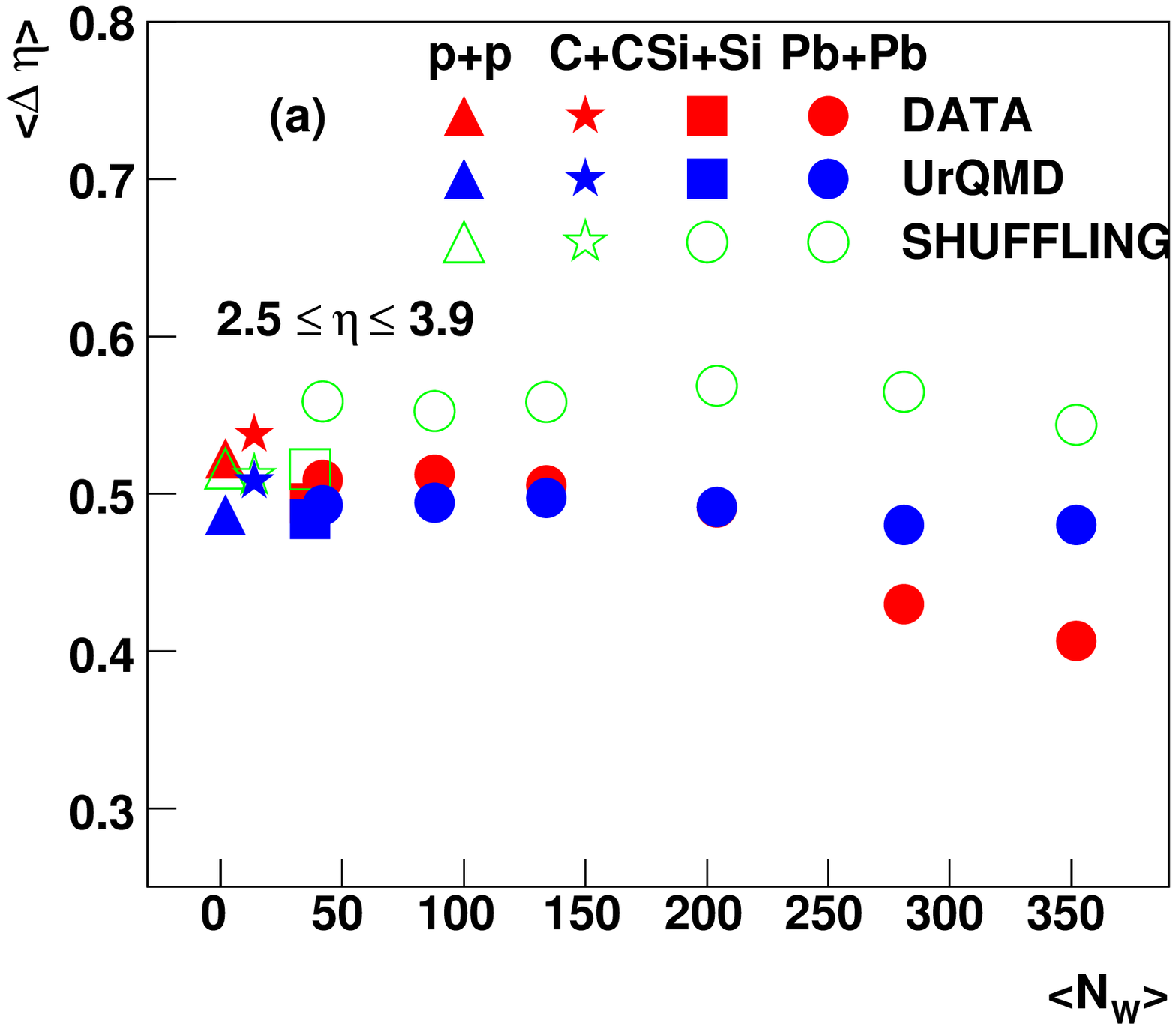}
\includegraphics[height=.22\textheight,width=.35\textwidth]{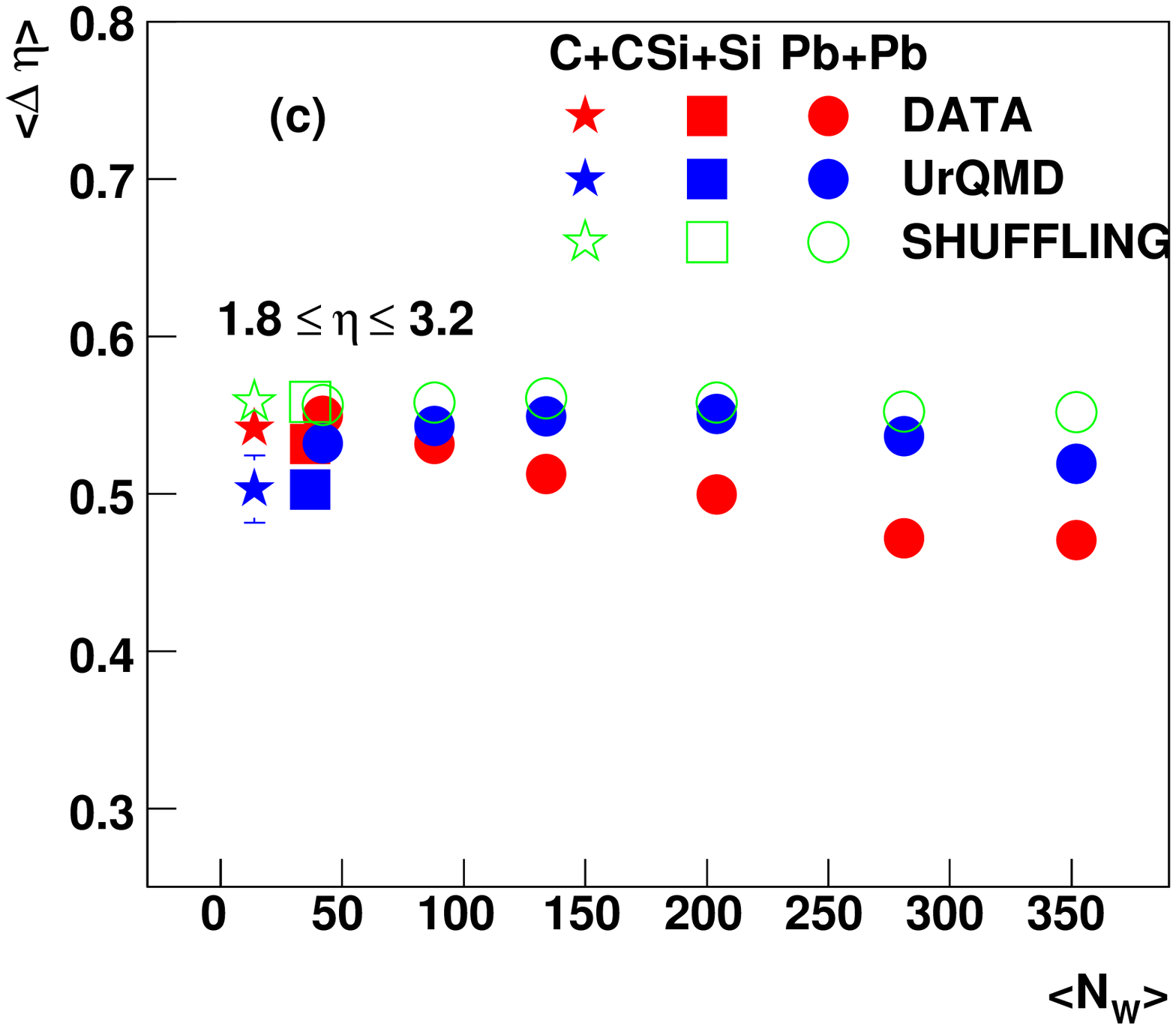}
\includegraphics[height=.22\textheight,width=.35\textwidth]{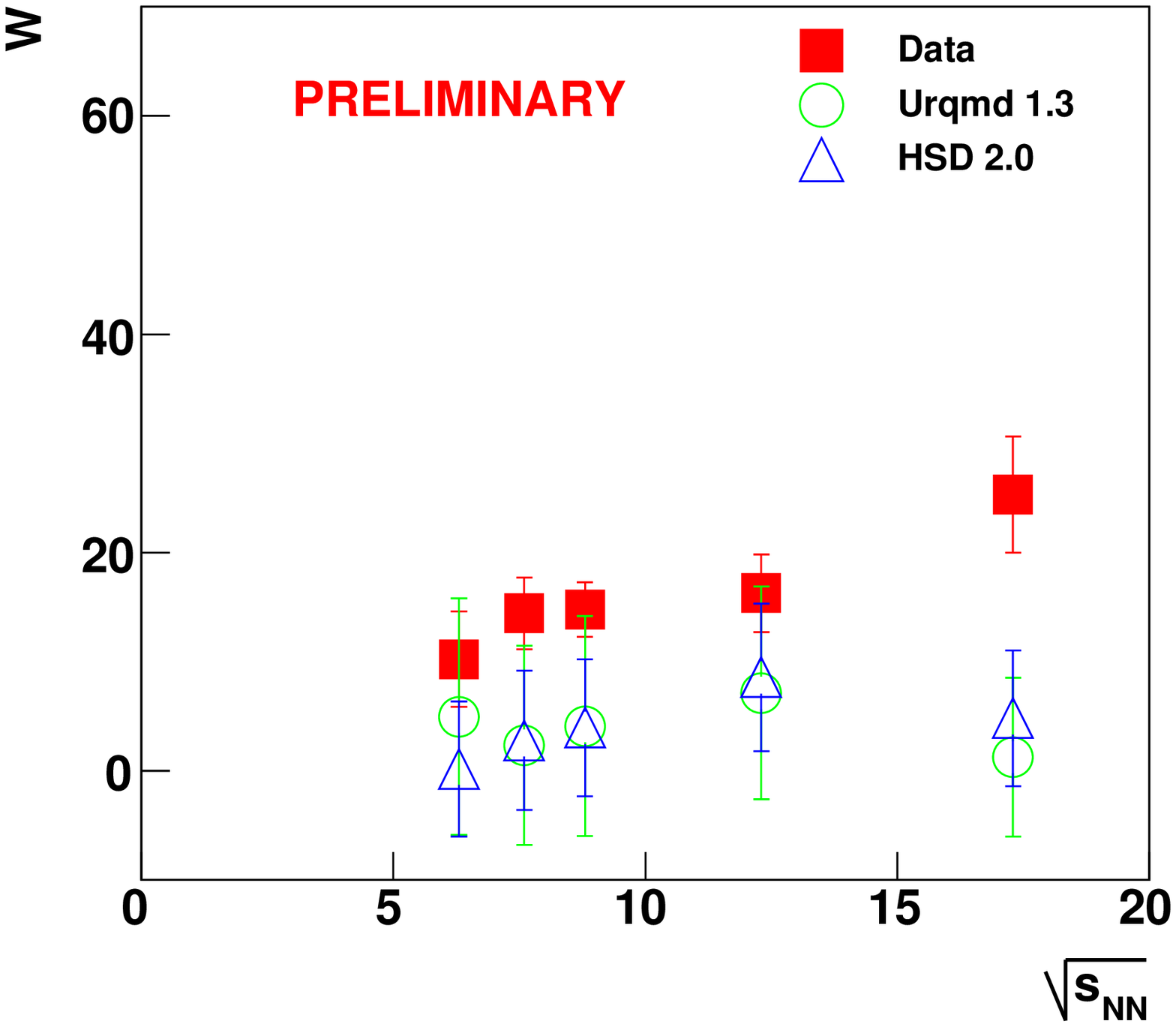}
\end{figure}

In addition, an attempt was made by NA49 to study the energy dependence of the Balance Function, by analyzing the most central Pb+Pb events throughout the whole available SPS energy range. The upper right plot of fig. \ref{FigNA49RapidityDependence} shows the dependence of the normalized W parameter, defined as $W = \frac{100 \cdot (\langle \Delta \eta \rangle_{shuffled} - \langle \Delta \eta \rangle_{data})}{\langle \Delta \eta \rangle_{shuffled}}$ \cite{NA49_BF}, on $\sqrt{s_{NN}}$. There is a first indication of an energy dependence for experimental points which is not reproduced by the microscopic models studied \cite{Urqmd,Hsd}. The lower right plot of fig. \ref{FigNA49RapidityDependence} shows the evolution of the W parameter from SPS to RHIC energies.

\begin{figure}
\includegraphics[height=.22\textheight,width=.35\textwidth]{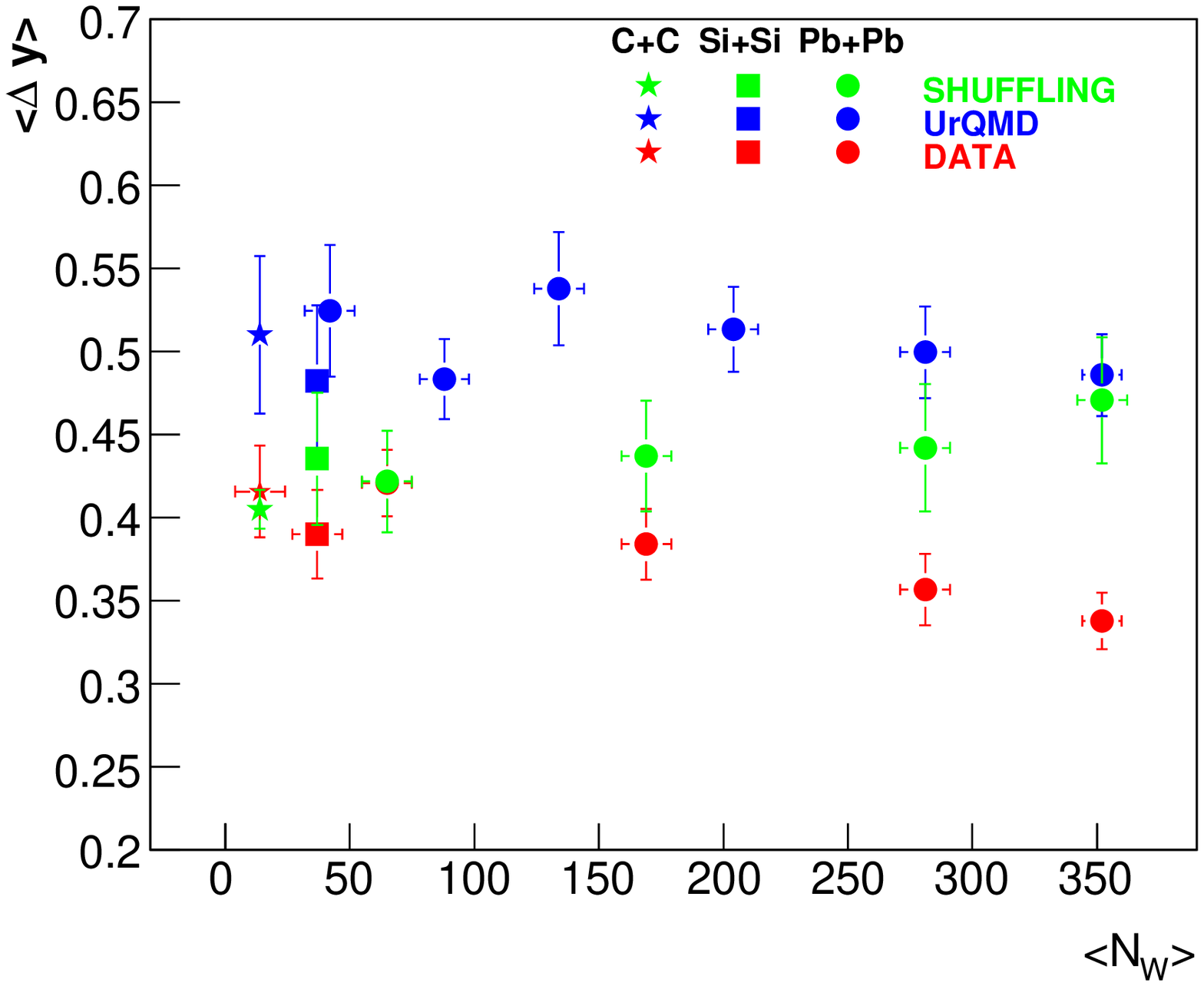}
\includegraphics[height=.22\textheight,width=.35\textwidth]{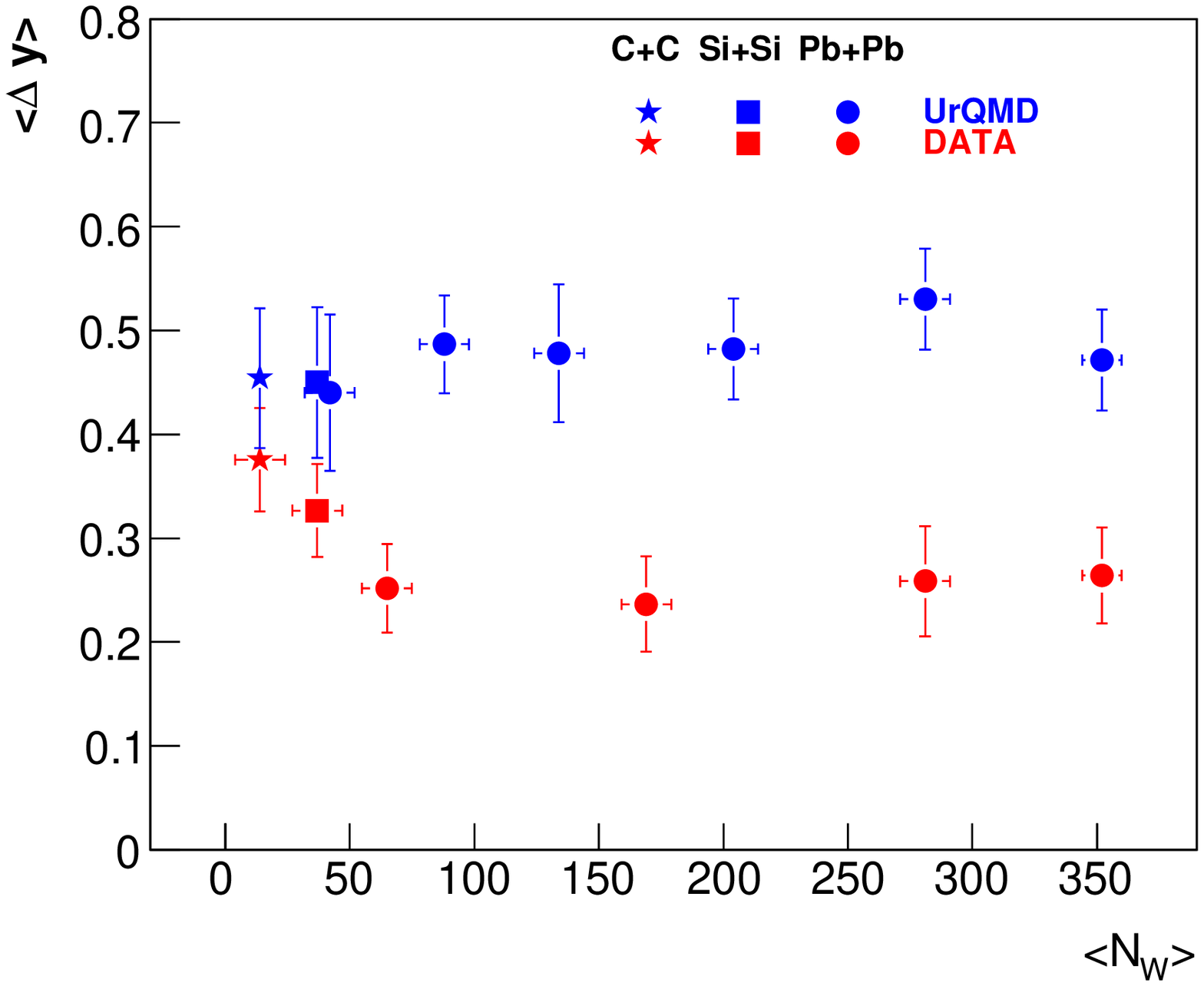}
\caption{The dependence of the width of the BF for identified pion (left plot) and kaon (right plot) pairs on $\langle N_{W} \rangle$ for A+A collisions at the highest SPS energy.}
\label{FigNA49BFPID}
\end{figure}

Finally, we studied the rapidity correlations of identified pion and kaon pairs, by selecting them using the dE/dx information from the TPCs, for the highest SPS energy and the corresponding results are summarized in fig. \ref{FigNA49BFPID}. The main conclusion is that there is a narrowing of the BF's width with centrality for the experimental data of pion pairs but not for kaon pairs. A similar behavior has also been reported by STAR \cite{STAR_BF}.

\begin{figure}
\includegraphics[height=.22\textheight,width=.35\textwidth]{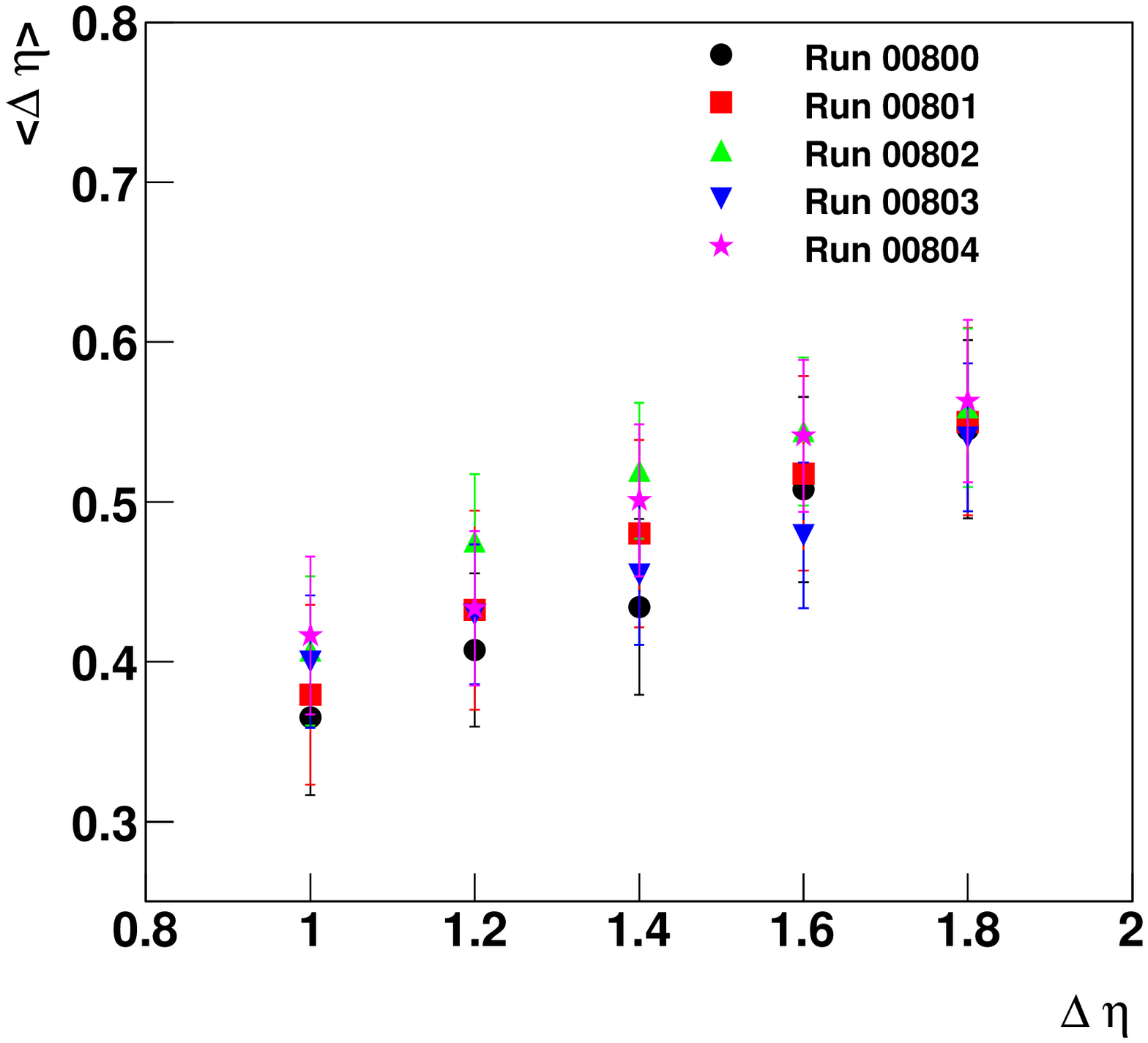}
\includegraphics[height=.22\textheight,width=.35\textwidth]{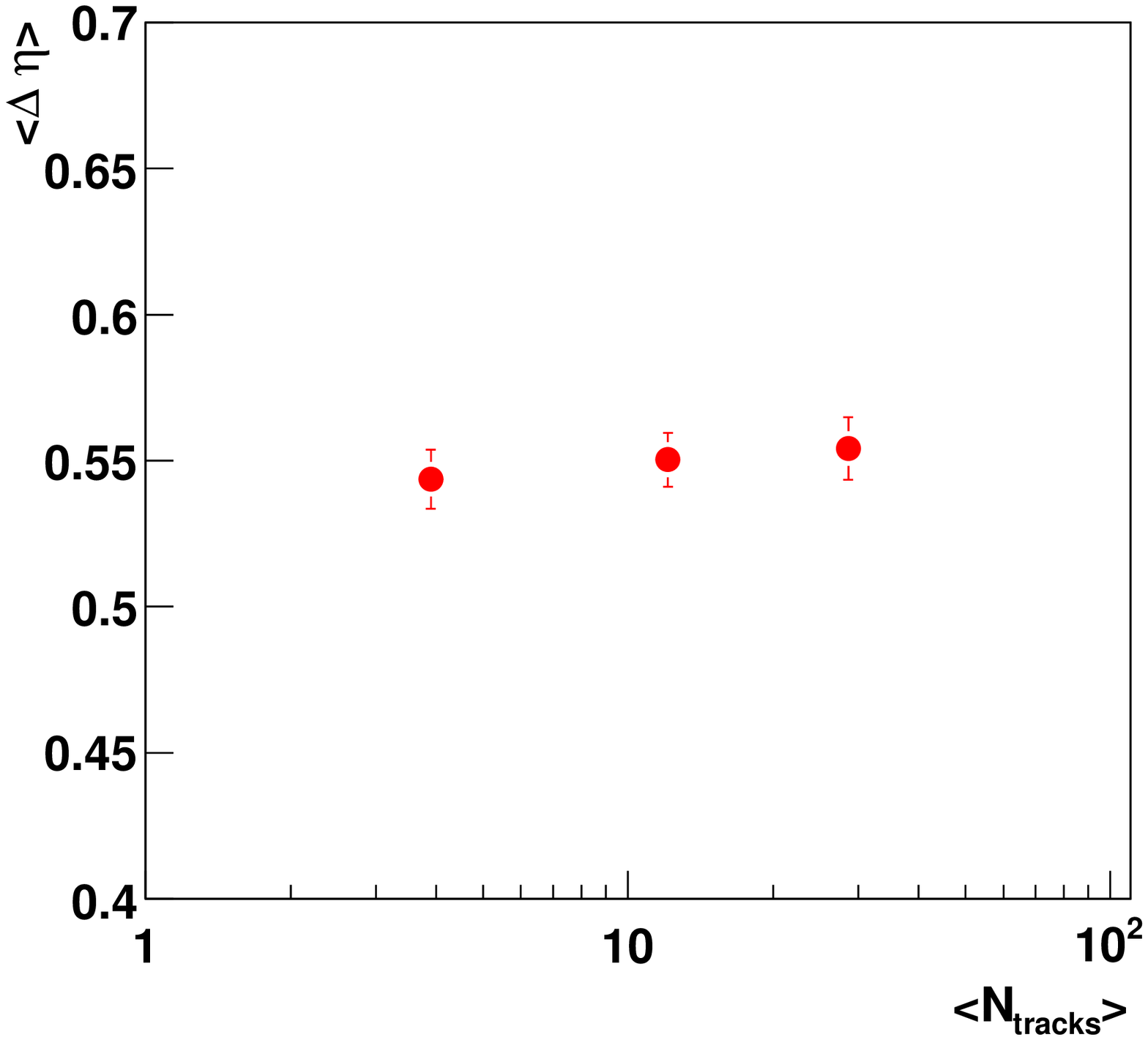}
\caption{The dependence of the BF's width on the analyzed interval (left plot) and on the mean number of tracks (right plot) for PYTHIA p+p events at $\sqrt{s} = 14$ TeV, using the framework of the ALICE experiment.}
\label{FigALICEBF}
\end{figure}


\vspace{0.1 cm}

The method was also extended to LHC energies by studying p+p PYTHIA events at $\sqrt{s} = 14$ TeV, generated and analyzed with the software framework of the ALICE experiment \cite{ALICEPPR}. Fig. \ref{FigALICEBF} shows the width of the BF for non identified particles as a function of the analyzed pseudo-rapidity interval (left plot) and of the mean number of tracks (right plot). A linear increase of the width on the analyzed interval is observed while no dependence on the mean multiplicity is found. 


\vspace{0.1 cm}

In summary, we presented the latest experimental results of the BF obtained by the NA49 collaboration. The rapidity dependence study revealed that the narrowing of the BF with centrality is restricted to the mid-rapidity region. In addition, the energy scan showed a first indication of a dependence of the normalized parameter W on $\sqrt{s_{NN}}$. The investigation of the rapidity correlations for identified pion and kaon pairs showed that we observe a narrowing of the BF's width with centrality for pions but not for kaons. Finally, we showed that the method has been extended to LHC energies by studying the hadron correlations for events simulated within the ALICE environment. 




\vspace{-0.5 cm}

\bibliographystyle{aipproc}

\end{document}